\documentclass[amsmath,amssymb,aps,floats,prl,twocolumn,showpacs,superscriptaddress,longbibliography]{revtex4-2}   
\usepackage{graphicx}  
\usepackage{dcolumn}   
\usepackage{bm}        
\usepackage{amssymb}   
\usepackage{xcolor}
\usepackage{amsmath}
\usepackage{soul}

\begin{document}

\title{\textcolor{black}{Terahertz displacive excitation of a coherent Raman-active phonon in V$_2$O$_3$}}



\author{Flavio Giorgianni}
\thanks{Corresponding author.\\ flavio.giorgianni@psi.ch}
\affiliation{Paul Scherrer Institute, 5232 Villigen-PSI, Switzerland}
\author{Mattia Udina}
\affiliation{ISC-CNR and Department of Physics, ''Sapienza'' University of Rome, P.le Aldo Moro 5, 00185 Rome, Italy}
\author{Tommaso Cea}
\affiliation{IMDEA Nanoscience, C/Farady 9, 28049 Madrid, Spain}
\author{Eugenio Paris}
\affiliation{SwissFel, Paul Scherrer Institute, 5232 Villigen-PSI, Switzerland}
\author{Marco Caputo}
\affiliation{Swiss Light Source, Paul Scherrer Institute, 5232 Villigen-PSI, Switzerland}
\author{Milan Radovic}
\affiliation{Swiss Light Source, Paul Scherrer Institute, 5232 Villigen-PSI, Switzerland}
\author{Larissa Boie}
\affiliation{Institute for Quantum Electronics, Physics Department, ETH Zurich, Auguste-Piccard-Hof 1, 8093 Zurich, Switzerland}
\author{Joe Sakai}
\affiliation{Institut Catal\`{a} de Nanoci\`{e}ncia i Nanotecnologia (ICN2), UAB Campus, ICN2 Building, 08193 Bellaterra, Spain}
\author{Christof W. Schneider}
\affiliation{Laboratory for Multiscale Materials Experiments, Paul Scherrer Institut, CH-5232 Villigen PSI, Switzerland}
\author{Steven Lee Johnson}
\affiliation{SwissFel, Paul Scherrer Institute, 5232 Villigen-PSI, Switzerland}
\affiliation{Institute for Quantum Electronics, Physics Department, ETH Zurich, Auguste-Piccard-Hof 1, 8093 Zurich, Switzerland}

\maketitle



\textbf{\abstractname}

Nonlinear processes involving frequency-mixing of light fields set the basis for ultrafast coherent spectroscopy of collective modes in solids. In certain semimetals and semiconductors, generation of coherent phonon modes can occur by a displacive force on the lattice at the difference-frequency mixing of a laser pulse excitation on the electronic system. Here, as a low-frequency counterpart of this process, we demonstrate that coherent phonon excitations can be induced by the sum-frequency components of an intense terahertz light field, \textcolor{black}{coupled to intraband electronic transitions}. This nonlinear \textcolor{black}{process leads to} charge-coupled coherent dynamics of Raman-active phonon modes in the strongly correlated metal V$_2$O$_3$. Our results show a new up-conversion pathway for the optical control of Raman-active modes in solids mediated by terahertz-driven electronic excitation.

\section*{Introduction}
Phononics has recently been proposed as a promising avenue towards a new generation of faster and more efficient technologies~\cite{phononics}. Phenomena associated with lattice dynamics are often analogous to well-known \textcolor{black}{effects} involving electrons, for example chirality and the phonon Hall effect~\cite{phononics1,phononics2,phononics3}. Moreover, control over coherent phonons offers potential new pathways to drive transitions between phases in condensed matter and to coherently manipulate the ground state of magnetically ordered phases~\cite{phasetrans,trsupercond,k3c60,magnetic,magnetic2}.  While coherent acoustic phonon modes are commonly generated by wide variety of mechanisms~\cite{acoustic}, higher-frequency coherent optical phonon modes are typically driven from interactions with intense sub-picosecond light pulses~\cite{phonon}.

In most cases, light \textcolor{black}{pulses} can drive optical phonons either directly \textit{via} \textcolor{black}{resonant} electric-dipole interaction~\cite{huber,resonantPh}, or indirectly \textit{via} interactions with electronic states or other phonon modes~\cite{magnetic,IRS,photoexcitation,THzupconversion}. \textcolor{black}{The latter case is a generalization of the \textcolor{black}{stimulated} Raman mechanism, in which the ultrashort light pulse couples to the electronic system resulting in a nonlinear force on the lattice that resonantly drives coherent phonon modes at the difference frequency mixing $\Omega=\omega_1-\omega_2$, with $\omega_1$ and $\omega_2$ taken for the pulse spectrum. This is commonly referred to in the literature as either Impulsive Stimulated Raman Scattering (ISRS), when the light field is non-resonantly coupled to the electric system (in optically transparent solids), or Displacive Excitation of Coherent Phonons (DECP), when the light pulse is resonant with intraband or interband electronic transitions (in opaque solids) ~\cite{ISRS,ISRS1,DECP1,DECP2}. Unlike ISRS, which involves non-resonant coupling to electronic states to produce impulsive excitation, in the DECP regime, coherent phonons are generated by a displacive force associated with a sudden change in carrier density or electronic temperature induced by the intense light pulse.} \textcolor{black}{A third generation mechanism is called Ionic Raman scattering (IRS) and involves a higher frequency phonon mode that serves as the intermediate~\cite{IRS}. In all these processes, the driving mechanism involves difference-frequency mixing, and requires the bandwidth of the driving light pulse to be comparable to or larger than the frequency of the optical phonon mode that is to be excited~\cite{bloembergen}.}

\textcolor{black}{As an alternative to these well-established routes for coherent phonon generation using stimulated Raman schemes, it is also possible to drive coherent Raman-active vibrational modes using sum-frequency excitation (SFE)~\cite{SFEP1,SFEP}. Recent experiments, indeed, have demonstrated both photonic and phononic SFE mechanisms~\cite{SFEP1,2dTHz,SFE2,SFE3,thzbi2se3}, which involve the absorption of two photons or two phonons with frequencies $\omega_1$ and $\omega_2$ to create one phonon in a Raman-active mode at frequency $\Omega=\omega_1+\omega_2$, as up-conversion counterparts of ISRS and IRS, respectively.}
 
Here, we demonstrate a novel scheme for \textcolor{black}{coherent generation} of Raman-active phonon modes based on terahertz-driven dispacive excitation as up-conversion counterpart of the DECP mechanism. An intense THz pulse, inducing strong transient \textcolor{black}{electronic excitation}, is found to generate large-amplitude coherent oscillations of Raman-active phonon modes in the metallic phase of V$_2$O$_3$ at twice the frequency of the driving field. \textcolor{black}{This nonlinear process can be described as a displacive excitation that involves the sum-frequency mixing component of the THz pulse coupled to the electronic system}. The temperature dependence of the generated coherent phonons supports the scenario in which the excitation process is strongly enhanced by intra-band electronic transitions driven by the intense THz field.

\section*{Results}
\subsection{Coherent phonon oscillations detected by THz pump - NIR probe spectroscopy}
In V$_2$O$_3$, strong electronic correlations cause a metal-insulator transition (MIT) at T$_{MIT} \simeq$ 150 K, which is accompanied by a first-order structural transition from corundum to monoclinic~\cite{v2o3,v2o31}. Recent experiments have shown that the coherent lattice displacement generated by perturbing the electronic distribution of strongly correlated metallic phase of V$_2$O$_3$  through ultrafast light excitation can drive both longitudinal acoustic waves and optical phonons~\cite{phononv2o3,phononv2o31,phononv2o32,phononv2o33,phononv2o34}. In these photoexcitation processes, the electron-phonon coupling plays a central role in the transient atomic displacement of the structure~\cite{correlation}.
\begin{figure*}
\vspace{0.53cm}
\includegraphics[scale=0.87]{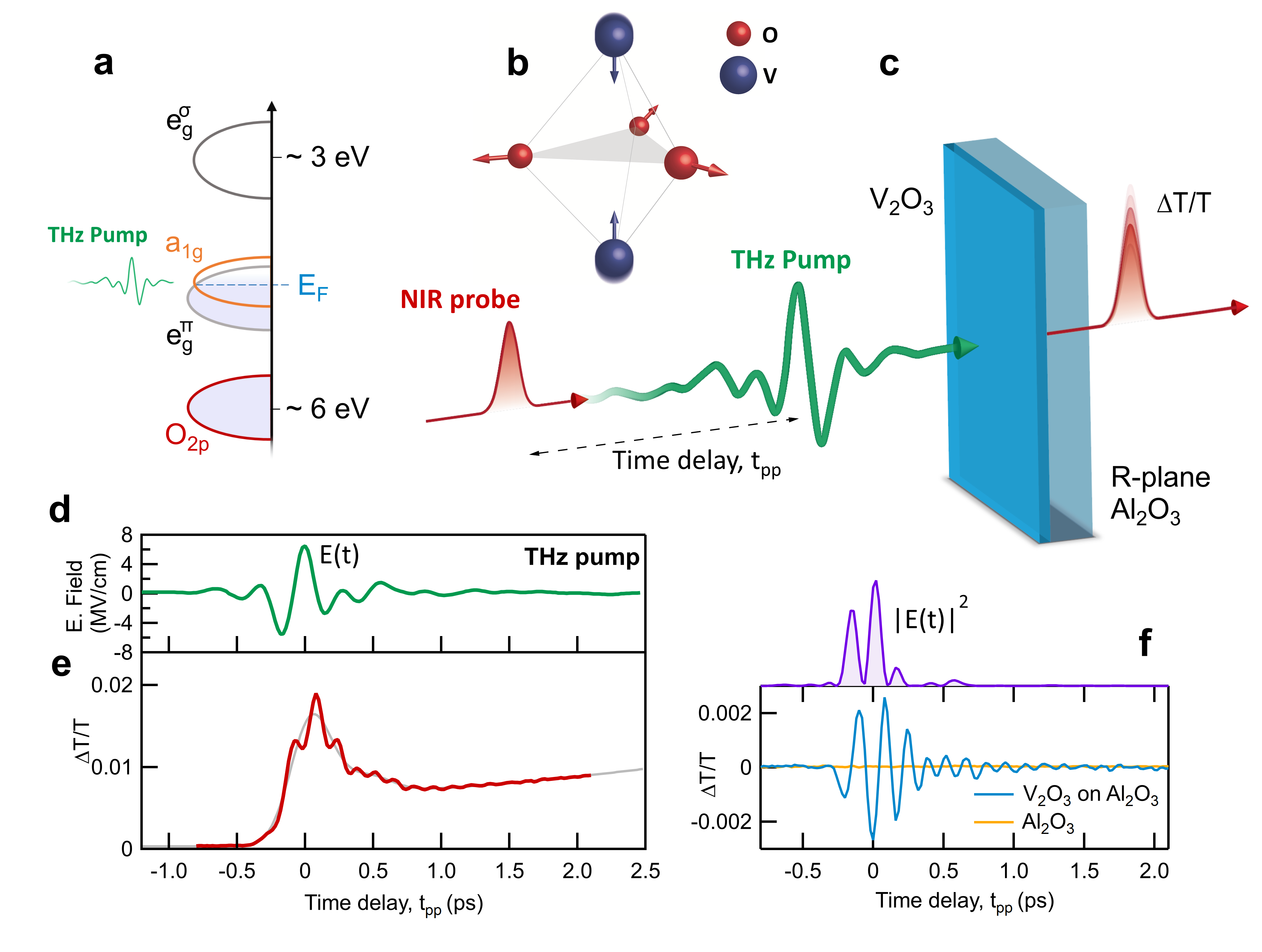}
\caption{\textcolor{black}{{\bf Coherent phonon generation by terahertz electronic excitations in V$_2$O$_3$.}} \textcolor{black}{{\bf a,} Energy level diagram for the paramagnetic metallic phase of V$_2$O$_3$ showing vanadium and oxygen energy levels. The Fermi energy E$_F$ crosses the partially filled $a_{1g}$ and $e_g^\pi$ bands formed by vanadium 3$d$ orbitals. O$_{2p}$ bands lie between 3.5 and 8.5 eV below E$_F$ while $e_g^{\sigma}$ is centered at $\sim$ 3 eV above E$_F$. Level diagrams adapted from Ref.~\cite{v2o31}}. {\bf b,} Phonon displacement vectors of the A$_{1g}$ phonon mode in V$_2$O$_3$. {\bf c,} Schematic of the setup for the THz pump-NIR probe experiment on V$_2$O$_3$ thin film on R-plane Al$_2$O$_3$ substrate to investigate the THz driven coherent lattice dynamics: intense THz pulse (green) excites the A$_{1g}$ phonon mode, whose coherent dynamics are probed by a temporally delayed ultrashort optical pulse (red). {\bf d,} Waveform of THz pump electric field. The peak electric field is 6.5 MV/cm. {\bf e,} Probe transmission modulation signal $\Delta T/T$ as a function of the pump-probe delay $t_{pp}$. The THz pump pulse induces coherent lattice oscillations related to the fully-symmetric A$_{1g}$ phonon at approximately twice the frequency of the driving electric field. The red curve represents the experimental data. Gray curve represents the slowly-varying background as determined by double-exponential relaxation fitting procedure (see Supplementary Note 4). {\bf f,} \textcolor{black}{Light-blue curve shows the oscillatory component of $\Delta T/T$ time-trace of V$_2$O$_3$ on R-plane Al$_2$O$_3$ after subtracting the background. Orange curve is the $\Delta T/T$ dynamics of the bare R-plane Al$_2$O$_3$ substrate measured in the same experimental conditions.} The purple curve shows the intensity profile $|E(t)|^2$ of the THz pump. Sample is at 300 K.}
\end{figure*}

\textcolor{black}{Fig. 1{\bf a} shows the schematic electronic energy levels for the metallic phase of V$_2$O$_3$. The Fermi energy crosses the partially filled $a_{1g}$ and $e_g^\pi$ bands, enabling the direct optical absorption of THz light by intraband electronic transitions~\cite{v2o31,v2o3}.} The $\Gamma$-point structural dynamics in this phase is largely dominated by a fully-symmetric  A$_{1g}$ optical phonon mode at $\Omega/2\pi=7.1$ THz, which corresponds to the optical bending mode of the V$_2$O$_2$ unit cell [see displacement vectors in Fig. 1{\bf b}~\cite{phononv2o32,ramanv2o3}]. Other Raman-active optical phonons in this frequency range, for example the E$_g$ phonon mode at $\sim$ 6 THz~\cite{ramanv2o3}, have a comparatively lower Raman cross-section, \textcolor{black}{as detailed in the Supplementary Note 3 and Supplementary Fig. 4.}

To study the possibilities for driving coherent structural dynamics in V$_2$O$_3$, we performed ultrafast THz pump-optical probe spectroscopy (scheme shown in Fig. 1{\bf c}). Intense quasi single-cycle THz pulses were focused onto the sample. The THz-induced response was monitored by measuring  the optical transmission modulation $\Delta T/T$ of a laser probe at 1.55 eV as a function of the time delay $t_{pp}$ with respect to the pump~\cite{birefringence,SFEP1}. The $\Delta T/T$ response is able to detect the coherent lattice oscillations from the fully-symmetric A$_{1g}$ \textit{via} time-dependent modulation of the diagonal elements of the linear optical susceptibility tensor (ST), while it is insensitive to the E$_g$ modes, which effect the probe polarization state through the modulation of the off-diagonal elements of the ST \textcolor{black}{(see detailed discussion in Supplementary Note 4)}.

The temporal waveform of the THz pump electric field $E(t)$ (see Fig. 1{\bf d}) was measured at the sample position using electro-optic sampling from a GaP crystal (see Supplementary Methods). The center frequency of the pump is $\nu_p \simeq$~3.5 THz.
The sample is a high-quality, 84-nm thick V$_2$O$_3$ thin film epitaxially grown on an \textcolor{black}{R-plane} (1 -1 0 2) $-$ Al$_2$O$_3$ substrate (see  Methods).
Figure 1{\bf e} shows the temporal evolution of the THz-induced transmission modulation $\Delta T/T$. The time origin ($t_{pp}=0$) is set to the delay where the optical probe interacts with the sample at the same time as does the peak of the square of the pump electric field $|E(t)|^2$. The observed THz-induced dynamics are dominated by coherent oscillations superimposed to a gradually increasing slowly-varying background. This long-lived step-like evolution of $\Delta T/T$ is attributed to incoherent dynamics of the electronic thermalization through electron-phonon scattering, as described in Refs.~\cite{phononv2o3,phononv2o32}.
Strikingly, the oscillating component of $\Delta T/T$ clearly displays oscillations with a period shorter than the one of the THz driving field. To better show this feature, in Fig. 1{\bf f} we compare the oscillatory component, obtained by subtracting the step-like background from $\Delta T/T$, with the intensity profile of the THz field $|E(t)|^2$. The fitting procedure for the background subtraction is reported in the Supplementary Note 4 and consists of Gaussian function to describe the THz excitation and a step-like function for the electronic relaxation dynamics as observed in Ref.s~\cite{phononv2o3,phononv2o32}. \textcolor{black}{To demonstrate that coherent oscillations originate from the V$_2$O$_3$ thin film, we conducted the same experiment on the bare R-plane Al$_2$O$_3$ substrate showing that there are neither oscillations nor a long-lived signal in the $\Delta T/T$ dynamics, see Fig. 1{\bf f} and Supplementary Note 2 for details.} Furthermore, from Fig. 1{\bf f} we note that the envelope function of the oscillatory response does not merely follow that of $|E(t)|^2$, but instead persists well beyond the times when the THz pulse intensity is negligible. This strongly suggests that the oscillations are indeed due to a coherent excitation of the A$_{1g}$ mode in V$_2$O$_3$ at its resonance frequency~\cite{ramanv2o3}.
\begin{figure*}
\includegraphics[scale=0.72]{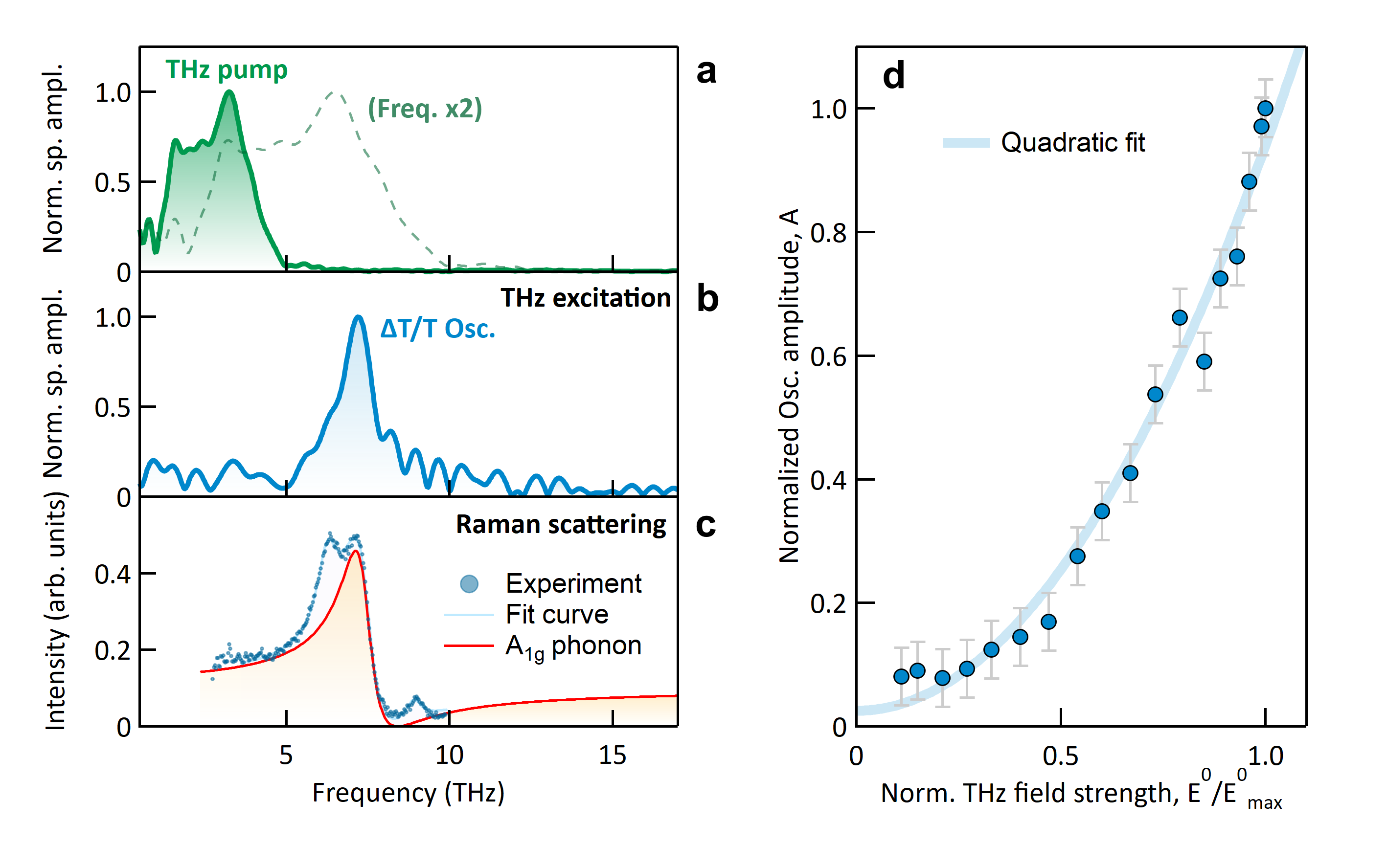}
\caption{ {\bf THz field-driven lattice phonon oscillations in frequency domain.} {\bf a,} THz spectrum of the pump pulse in Fig. 1{\bf d} (green solid curve) and same spectrum with 2x frequency (green dashed curve). {\bf b,} Fourier transform of the oscillatory component of $\Delta T/T$ for $t_{pp}>0.4$ ps (blue curve). {\bf c,} Raman spectroscopy of V$_2$O$_3$ thin film at 300 K. Light blue dots are experimental data and light solid curve is the fit curve, which contains different spectral components: two E$_g$ modes and the A$_{1g}$. The red curve is lineshape of the A$_{1g}$ phonon, which is peaked at 7.1 THz. Vertical offset of -0.5 has been added to the experimental data for the overlap with the A$_{1g}$ phonon lineshape. Data without offset is shown in Fig. S4 and discussed in Supplementary Note 3. {\bf d,} Amplitude of the coherent oscillations \textit{vs} the normalized electric field strength, $E_0$/$E_{peak}$, with $E_{peak}$=6.5 MV/cm. The amplitude scales quadratically with the pump electric field (the light blue solid line represents a quadratic fit).}
\end{figure*}
Such effect is clearly visible in the frequency domain where Fourier amplitude of the observed optical transmission oscillations show a narrow response where the pump spectrum has a negligible amplitude (Fig. 2{\bf a}-{\bf b}). The spectrum of the oscillatory component peaks at the A$_{1g}$ phonon frequency measured by Raman spectroscopy, which matches the double of the frequency of the driving field (Fig. 2{\bf c}). As one can see, the spectrum of the A$_{1g}$ Raman spectrum shows a narrower line shape than the frequency response of the oscillations of the transmission modulation (Fig. 2{\bf b}), because its spectral contents also depend on the frequency of the pump stimulus, which give rise to further broadening. This effect is also observed in V$_2$O$_3$ with conventional pump-probe optical spectroscopy~\cite{phononv2o33}. \textcolor{black}{Furthermore, to exclude the contribution of other modes to the observed dynamics, we repeated the experiment on a V$_2$O$_3$ thin film deposited on a C-plane (0001) Al$_2$O$_3$ substrate, whose orientation results in a diagonal susceptibility tensor allowing only for the fully-symmetric A$_{1g}$ modes to be detected (see Supplementary Note 3). The pump-probe dynamics of this sample also shows coherent oscillations peaked at the frequency of the A$_{1g}$ phonon mode (see Supplementary Fig. 5), further demonstrating that the phonon generated by the THz excitation is the A$_{1g}$ mode.}

Fig. 2{\bf d} shows the dependence of the oscillation amplitude \textit{A} on the electric field amplitude \textit{E$^0$} of the incident THz field. The dependence is well described by a quadratic trend \textit{A=a(E$^0$)$^2$} proving that the mode is nonlinearly coupled to the driving electric field. Moreover, as the spectral content of the pump pulse contains negligible intensity above 5 THz, we exclude a linear (dipolar) coupling of light to the lattice mode, which is in any case symmetry-forbidden since the mode is purely Raman active.

\subsection{\textcolor{black}{THz displacive excitation of phonons}}
Next, we discuss the nonlinear excitation mechanism of the vibrational modes by THz driven electronic excitation.
We propose that the interaction of the THz-frequency pulse with the metallic phase of V$_2$O$_3$ first leads to the excitation of a time-dependent electronic distribution $\eta(t)$ in the conduction bands. The time-domain dynamics of $\eta(t)$ by THz excitation are described by:
\begin{equation}
\begin{split}
\eta(t)=\kappa \int_{0}^{\infty}E^2(t-\tau) \left(e^{-\beta\tau}+c \right) d\tau,
\end{split} 
\end{equation}
where $E(t)$ is the electric field temporal waveform of the THz pump pulse, $1/\beta$ is the fast electronic relaxation rate after the THz excitation and $c$ accounts for long-lived thermalization dynamics that last longer than 2 ps as also observed in V$_2$O$_3$ by optical pump-probe spectroscopy~\cite{ramanv2o3,phononv2o33}
The excited electronic states interact with the A$_{1g}$ lattice mode \textit{via} electron-phonon coupling and, in a semi-classical picture, they exert an effective force on the A$_{1g}$ mode that has a time-dependence given by the intensity of the THz light field.  
\begin{figure*}
\includegraphics[scale=0.52]{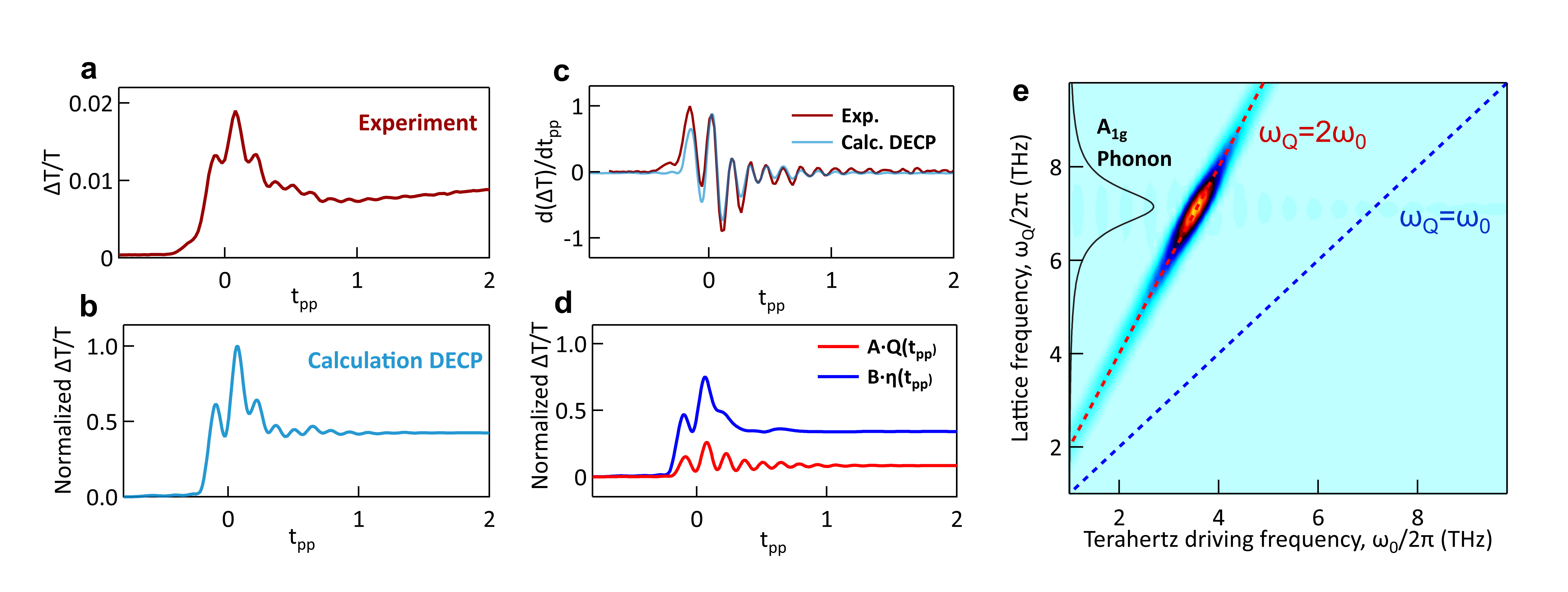}
\caption{{\bf Modelling of terahertz \textcolor{black}{displacive} excitation of phonons.} {\bf a,} Experimental data at 300 K and {\bf b,} numerical calculation using Eq. 3 using phononic and electronic parameters of V$_2$O$_3$ at 300 K. {\bf c,} Comparison of the time derivative of the dynamics: experimental data (red curve) and numerical calculation (light blue curve). {\bf d,} Calculated electronic (blue curve) and lattice (red curve) contributions to the $\Delta T/T$ dynamics from Eq. 3. The electron dynamics show no signature of long lived oscillatory components. {\bf e,} Time-derivative of the lattice dynamics $Q(t)$ in frequency domain calculated by DECP model [Eq. (2)] as a function of the center frequency $\omega_0/2\pi$ of the driving THz field. Temporal derivative is shown to emphasize the oscillatory component over the slowly-varying background. The phonon amplitude is maximum when $\Omega=2\omega_0$ as described by THz SFE. The black curve is the A$_{1g}$ phonon lineshape as a function of the lattice frequency $\omega_Q/2\pi$.}
\end{figure*}
According to general theory of  DECP (see Ref.~\cite{DECP1}), the \textcolor{black}{effective} force acting on the lattice \textcolor{black}{associated with the} electronic perturbation \textcolor{black}{due to the} intense time-varying electric field $E(t)$ \textcolor{black}{leads to} the following equation of motion \textcolor{black}{for the phonon displacement field}:
\begin{equation}
\ddot{Q}+2\gamma \dot{Q}+\Omega^2Q=\alpha \eta(t) ,
\end{equation}
where $Q$ is the normal coordinate of the A$_{1g}$ phonon, $\Omega/2\pi$ the phonon eigenfrequency, $\gamma/2\pi$ the phonon damping rate~\cite{ramanv2o3,phononv2o33} and $\alpha$ the coupling constant.
We model the THz-induced transmission modulation as the sum of the lattice and electronic contributions [Eq.s (1) and (2)], as described in Ref.~\cite{DECP1}:
\begin{equation}
\Delta T/T(t_{pp})=AQ(t_{pp})+B\eta(t_{pp}), 
\end{equation}
with A and B multiplying constants.

Under this assumption, from Eq. 3 we calculated numerically the time-evolution of the transmission modulation dynamics $\Delta T/T(t_{pp})$ by using the experimental THz driving field shown in Fig. 1{\bf c} and the following A$_{1g}$ phonon parameters $\Omega/(2\pi)=7.1$ THz, $\gamma/(2\pi)=$ 0.5 THz obtained by Raman spectroscopy (see Supplementary Note 3 and Supplementary Table 2). For the fast electronic relaxation dynamics after the THz excitation we set $1/\beta$= 80 fs in agreement with Ref. ~\cite{phononv2o32}; For the electron-phonon relaxation dynamics we set $c$ = 0.2 and $B/A$ = 0.002 to fit the amplitude of the experimental data. As expected from the numerical result of Eq. 3, the calculated dynamics by DECP model exhibits coherent lattice oscillations at twice the frequency of the electric field (Fig. 3{\bf a} and Fig. 3{\bf b}) and a step-like dynamics in agreement with the experiment, as also shown by the comparison of the corresponding time derivatives, see Fig. 3{\bf c}. Fig. 3{\bf d} shows both the calculated electronic and lattice components of the dynamics from Eq. 3. As observed, the electronic dynamics has no long-lasting oscillating features.

Furthermore, we compute the lattice dynamics $Q(t)$ by DECP model (Eq. 2) as a function the driving frequency of THz pump. The numerical calculation of $Q(t)$ is performed for multiple Gaussian THz pulses with a full width half maximum of 0.5 THz and center frequency $\omega_0/2\pi$, spanning from 1 to 10 THz \textcolor{black}{with a frequency step of 0.125 THz}. In Fig. 3{\bf e}, we show the temporal derivative of $Q(t)$ in frequency domain, which shows a maximum when THz driving frequency $\omega_0/2\pi$ is half of the phonon frequency $\omega_Q/2\pi$ as described by SFE, while no excitation is observed in resonant condition $\omega_Q=\omega_0/2\pi$.

\subsection{THz-induced dynamics across the temperature driven insulator-to-metal transition}
We now consider the temperature dependence of the oscillatory response.  Fig. 4{\bf a} shows the phase diagram of $($V$_{1-x}$M$_x)_2$O$_3$, where the dopant M is either Cr or Ti.  We drive the temperature across the metal-to-insulator transition at $x=0$.
The temperature evolution of $\Delta T/T(t_{pp})$ across the MIT is shown in Fig. 4{\bf b} for various temperatures between 77 K and 500 K. In the insulating phase (T$<$T$_{MIT}$) the transmission modulation displays a negative step-like component (see Fig. 4{\bf c}) due to a partial formation of the metallic phase by the intense THz field~\cite{thzv2o3,thzvo2}. However,  the amplitude of coherent lattice vibrations is strongly reduced in the insulating phase (see Fig. 4{\bf d} and Supplementary Fig. 7). In contrast, as the temperature approaches the MIT, we observe an emerging oscillatory signal at 2$\omega_p/2\pi$, which denotes that the phonon excitation mechanism becomes efficient when the THz pump is resonant with the conduction states. By further increasing the temperature, the transmission modulation response rapidly decreases above the coherence temperature (T$_{ch}$~$\sim$~400 K), see Fig. 4{\bf d}. This may be due to the suppression of the optical conductivity by means of the enhanced scattering rate of the electrons in the correlated electronic phase of V$_2$O$_3$ at T$_{MIT}<$T$<$T$_{ch}$~\cite{coher,phasediagram}.
Remarkably, the amplitude of the oscillations versus the temperature strongly recalls the THz optical conductivity, as shown in the Fig. 4{\bf d}-{\bf e} At present, an accurate model describing the transmission modulation response as a function of the temperature is still lacking, as it must account for the evolution of both the electronic phase and the lattice structure across the different phases. We can, however, qualitatively understand the data shown in Fig. 4{\bf d}-{\bf e} by noting that the \textcolor{black}{THz generated} displacive force on the lattice coordinate (see Eq. 1 and Ref.~\cite{DECP1}) is proportional to the deposited THz energy which is, in turn, proportional to the real part of the optical conductivity $\sigma_1$ according to the Joule heating model: $\int \sigma_1 E^2(t) dt$, being the imaginary part $\sigma_2$ of metallic phase negligible in the THz range~\cite{sigma2V2O3}. \textcolor{black}{Therefore the temperature trend of the phonon amplitude and of $\sigma_1$ further supports the interpretation that the underlying generation mechanism is displacive rather than impulsive in nature.}
\begin{figure}
\includegraphics[scale=0.31]{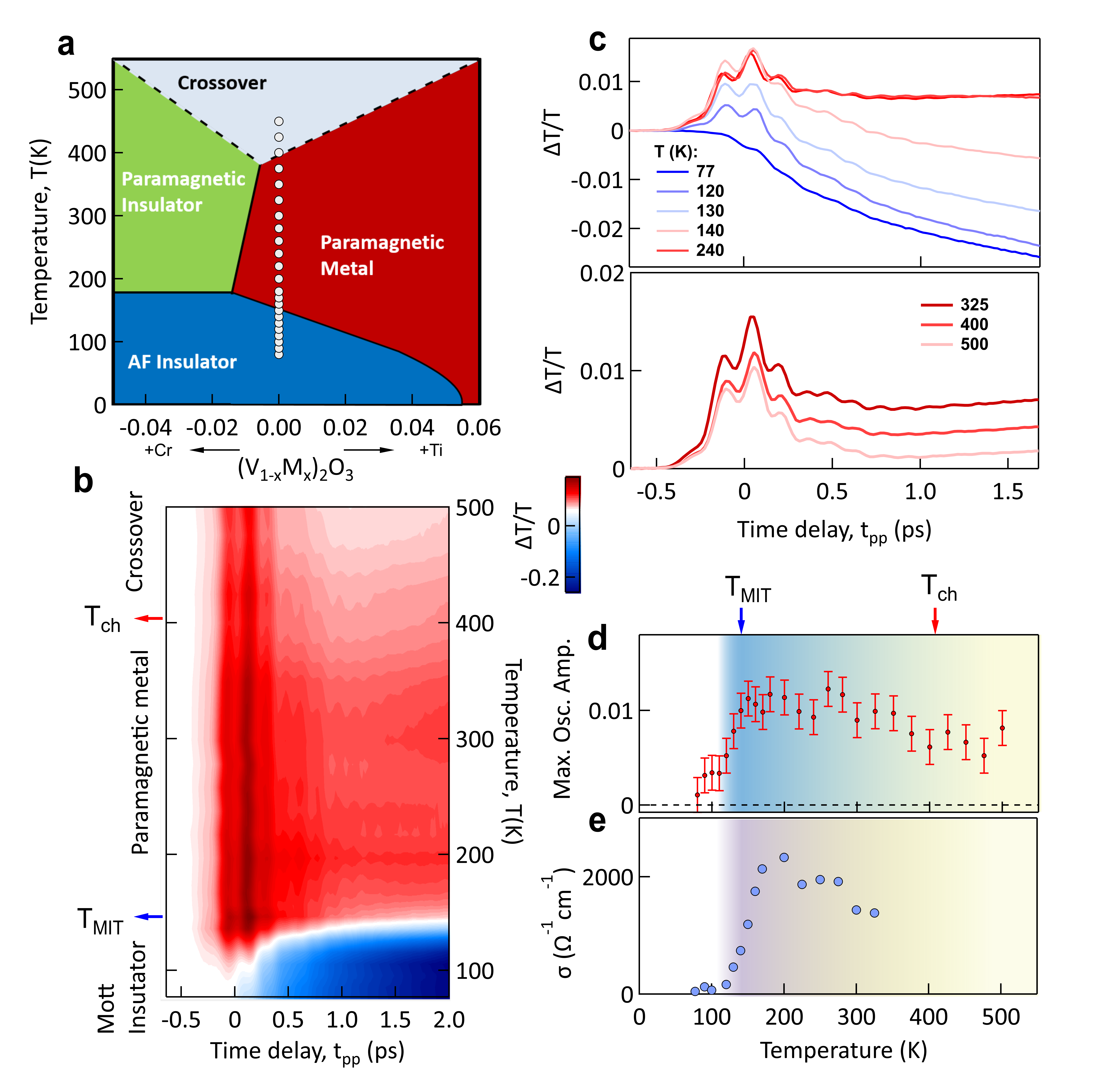}
\caption{{\bf Temperature evolution of the sum-frequency excitation across the insulator to metal transition.} {\bf a,} Temperature vs doping phase diagram of V$_2$O$_3$ according to Ref.~\cite{phasediagram}. The empty circles denote the values of the temperature measured in the present work. {\bf b,} Temperature dependence of the THz pump induced optical transmission modulation in V$_2$O$_3$. When the THz pulse interacts with the insulating Mott phase, no prominent signatures of lattice oscillation by SFE are observed. {\bf c,} Pump-probe traces at selected temperatures. {\bf d}-{\bf e,} Temperature dependence of the oscillations amplitude {\bf d,} and the THz optical conductivity {\bf e} of V$_2$O$_3$. THz optical conductivity $\sigma_1$ of the V$_2$O$_3$ film as a function of temperature was measured using Fourier-transform infrared spectroscopy in the THz range (see Supplementary Note 1).}
\end{figure}

\section*{Discussion}
We propose the following scenario to describe the mechanism \textcolor{black}{for the nonlinear generation} of a coherent A$_{1g}$ phonon mode in V$_2$O$_3$. \textcolor{black}{First, the incident THz field drives primarily intraband transitions in the a$_{1g}$ electronic band of 3$d$ vanadium orbitals having the dominant contribution in the Drude spectral weight compared to the ones from the e$_g^\pi$ band~\cite{v2o31}.} These newly populated states exert an effective force on the ions, driving a coherent A$_{1g}$ symmetry mode. Okamoto et al.~\cite{ramanv2o3} have shown that Raman intensity of A$_{1g}$ phonon modes are \textcolor{black}{dominated} by the electronic occupancy of a$_{1g}$ band mediated by electron-phonon coupling ~\cite{ramanv2o3,zhao1,zhao2}. Phenomenologically, the sum frequency excitation of the phonon mode in metallic V$_2$O$_3$ can be explained because the intense THz transient can temporally control the electronic distribution by intraband transitions in the a$_{1g}$ bands, proportional to the THz intensity, leading to the coherent up-conversion excitation of the A$_{1g}$ phonon mode. It is noteworthy that this process does not take place in insulating \textcolor{black}{phase of} V$_2$O$_3$, where the THz pulse is not coupled to a$_{1g}$ electronic states, while non-resonant SFE is not observed because the Raman cross-section is lower than that of diamond crystal~\cite{SFEP1}. Finally, we note that time-resolved optical spectroscopy is not \textit{a priori} able to determine the coefficients in Eq. 3; a quantitative disentangling of the electronic and lattice driven responses requires additional information that could be provided by future ultrafast x-ray or electron diffraction experiments to determine the magnitude of the coherent lattice displacements.
An interesting connection exists between our observation of THz light-induced \textcolor{black}{coherent phonon generation} via stimulated Raman scattering from conductive electrons and the phenomenon of electrostriction in metals for DC electric fields.  In electrostriction mechanism, an intense current in metals creates a strain field, proportional to the square of the electric field ($E^2$), due to interaction of conduction electron states with acoustic phonon modes~\cite{electrosctiction}.  Electrostriction effects at THz frequencies have been reported in metallic films, where the strain waves driven by the intense THz field can generate mechanical fractures~\cite{thzelectrosctiction,thzv2o3}.

Our observation of the coherent generation of Raman active phonons by THz excitation in a metallic system provides an interesting counterpart to ISRS- and DECP-based schemes in the infrared and visible ranges, using photons with much lower energy than the excited mode. This has the potential advantage of limiting the inevitable heating of the electronic system that necessarily occurs as a side effect of any scheme using light to coherently control the structure of metallic systems. This displacive excitation scheme may have potential advantages for applying coherent control methods to target other \textcolor{black}{lattice-coupled collective modes, including magnons, for which sum-frequency inonic excitation has been recently  discussed theoretically~\cite{SFEM},} and provide new insights about the interplay between the lattice and other coexisting ordered phases.

\vspace{1cm}
\centerline{\textbf{Sample characterization and growth}}
\noindent The growth technique, diffraction analysis and transport measurements of the sample used in the experiment are reported in Ref. ~\cite{jsakai}. Specifically the sample (84-nm-thin epitaxial film of V$_2$O$_3$) has been grown on an R-plane sapphire substrate by pulsed laser deposition with a fluence of 1 J/cm$^{2}$. Raman spectroscopy of the V$_2$O$_3$ film at 300 K (fig. 3{\bf c} in the main manuscript) has been carried out by using a LabRAM Series Raman Microscope (Horiba Jobin Yvon) equipped with a He-Ne excitation laser ($\lambda$= 632.8 nm).





\pagebreak
\onecolumngrid

\setcounter{figure}{0}
\renewcommand{\thefigure}{S\arabic{figure}}

\renewcommand{\thetable}{S\arabic{table}}

\setcounter{equation}{0}
\renewcommand{\theequation}{S\arabic{equation}}

\renewcommand{\thesubsection}{S\arabic{subsection}}
\renewcommand{\appendixname}{}

\begin{center}
\textcolor{white}{}\\
\vspace{0.5cm}
\large\textbf{Supplementary Information}
\end{center}
\vspace{0.cm}

\section*{Supplementary Methods. Experimental setup and THz pump field characterization}

The experimental setup is based on a 100 Hz, 20 mJ, 50 fs Ti:sapphire laser system coupled with an optical parametric amplifier (OPA). The OPA output energy (signal+idler) was 6.2 mJ/pulse. The OPA signal (1.5 $\mu$m wavelength) was used for the generation of intense single-cycle THz pump pulses by optical rectification in a large-size DSTMS crystal from Rainbow Photonics~\cite{ref1}. The OPA signal beam energy was reduced to 2 mJ/pulse by using a half wave-plate and NIR polarizer to prevent sample damage by high THz electric field. The organic crystal has a thickness of 0.41 mm and a clear aperture of 7 mm, while the OPA signal beam diameter is $\sim$9 mm at the crystal surface. Metallic multi-mesh filters were inserted into the THz optical path to block the residual OPA beam after the optical rectification process and to shape the THz pump spectrum. Specifically, in the experiment we used two 20 THz low-pass filters (LPFs) and one 6 THz LPF (from QMC Ltd). The measured NIR extinction ratio of the THz filters is $7\cdot10^7$.

Furthermore, the absence of the OPA beam at the sample position was experimentally checked by means of a 4 mm-thick CaF$_2$ window inserted into the beam path: by using the CaF$_2$ window, which transmits $\sim 95\%$ of infrared light but it is opaque to the THz radiation, no signal related to NIR radiation was detected on both a microbolometric THz camera (NEC IRV-T0830) and an infrared camera (WinCamD-UCD12).

The THz pump beam was focused on the sample by means of three off-axis parabolic mirrors with reflected focal length (RFL) of 0.5'', 4'' and 2'', respectively (see Fig.\ S1{\bf a}). The resulting THz beam is tightly focused onto the sample. To vary the THz electric field strength on the sample, we used a pair of wire-grid THz polarizers: the first one (TP1) at a variable angle $\theta$, the second one (TP2) fixed to select the vertical polarization component (see Fig.\ S1{\bf a}).

The THz transverse profile was measured by using a THz microbolometric camera (NEC IRV-T0830) while the THz pump temporal profile was characterized at the sample position by electro-optic sampling (EOS) method using a (110) GaP crystal with a thickness of 200 $\mu$m. The THz energy/pulse was measured by a Gentec THZ12D-3S-VP-D0 energy meter.

For the THz pump pulse in Fig.\ 1 of the main manuscript, we measured the following parameters: THz energy/pulse 6.83 $\mu$J, focus spot size (1/e$^2$ level): 212 $\mu$m from Gaussian fit of the beam profile (see Fig.\ S1{\bf c}-{\bf d}).

The measured electric field strength E$_{THz}$= 6.5 MV/cm was evaluated using the following formula~\cite{ref2}:
\begin{equation}
E_{THz}=\sqrt{\frac{z_0W}{\pi w^2 \int_{-\infty}^{\infty}g^2(t)dt}},
\label{eqS1}
\end{equation}
where $z_0=377$ $\Omega$ is the free space impedance, $W$ is the THz energy/pulse, $w$ is the beam waist and $g^2 (t)$ is the temporal intensity profile of the THz pulse with peak value normalized to 1, (see Fig.\ S1{\bf b}). The temporal profile $g(t)$ was measured by EOS. We obtain  $\int_{-\infty}^{\infty}g^2 (t)dt= 0.172$ ps.
\begin{figure}[h!]
\centering
\includegraphics[scale=0.650]{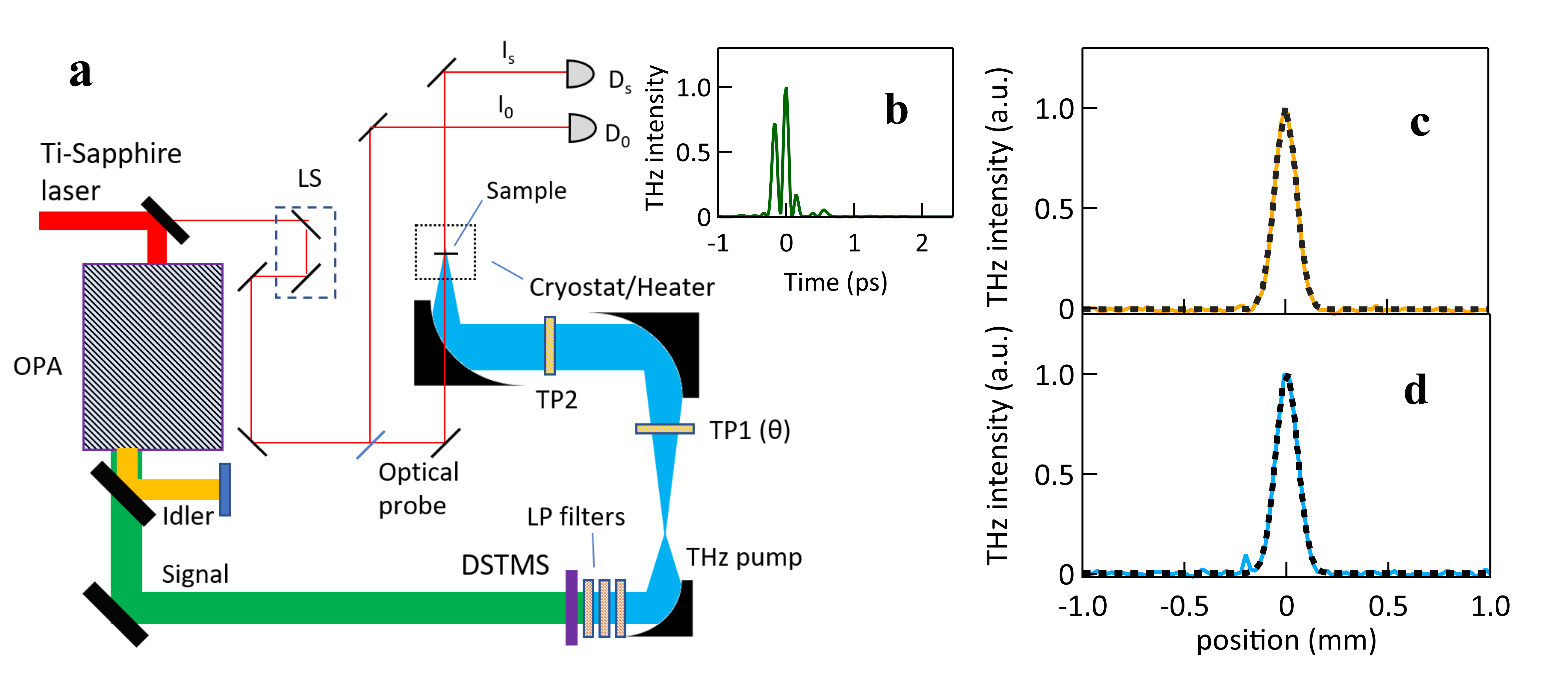}
\caption{\textbf{Experimental setup and THz pump characterization}. {\bf a}) THz pump-800 nm probe setup. LS: linear stage; TP1: THz Polarizer 1; TP2: THz Polarizer 2; LP (low-pass) filters; D$_s$ and D$_0$ are detectors to measure the transmitted probe beam from the sample and the reference beam, respectively. {\bf b}) Intensity profile $|E(t)|^2$ of THz pump. {\bf c}) Vertical and {\bf d}) horizontal THz beam transverse profile of THz pump at the sample position. Yellow and blue solid curves are the experimental data measured with a THz camera while black dotted lines are the Gaussian fits.}
\end{figure}

As a crosscheck of the THz pump electric field strength, we performed electro-optic sampling at the sample position in the same configuration by using a 50 $\mu$m-thick GaP crystal. In order to prevent optical over-rotation effect by the strong field, the THz beam was attenuated by means of a calibrated THz ND filter (from Tydex) with 10 $\%$ of transmission and by rotating TP1 to $\theta=65^{\circ}$ (measured overall THz transmission amplitude = 0.016$\pm$0.003). The uncertainty is dominated by the angle and a possible contribution of the residual background.

We calculated the peak electric field with the formula~\cite{ref3}:
\begin{equation}
\frac{S_2-S_1}{S_2+S_1}=\frac{2\pi}{\lambda_0}  n_0^3 r_{41} t_{GaP} E_{THz} d,
\label{eqS2}
\end{equation}
where $n_0 = 3.11$ is the index of refraction of GaP at $\lambda_0 = 800$ nm, $r_{41}$ = 0.97 pm/V is the electro-optical coefficient of GaP and d = 50 $\mu$m is the crystal thickness. $t_{GaP}=0.46$ is the Fresnel transmission. Taking into account the attenuation of the THz ND filter and THz polarizers, we calculate a peak THz electric field of 6.2 MV/cm. Table S1 reports the comparison between the two methods.

The THz pump-induced transmission modulation of the V$_2$O$_3$ samples is measured by a temporally delayed 800-nm probe pulse derived from the Ti:sapphire laser beam split before the OPA (see optical scheme in Fig.\ S1{\bf a}). Prior to the interaction with the sample, a beam splitter separated a portion of the optical probe beam (I$_0$) as reference for the differential transmission modulation. The 800-nm probe has a spot size of $\sim 45 \mu$m (1/e$^2$) width at the sample position measured by using a NIR camera (WinCamD-UCD12). This spot size is less than half of that of the THz pump at the sample position. The probe beam is linearly polarized and vertically oriented. The probe beam undergoes a change in transmission induced by the co-propagating THz beam in the sample.

\begin{table}[h]
\centering
\renewcommand{\arraystretch}{2.5}
\begin{tabular}{| c | c | c |}
\hline
& From THz Energy, Eq.\ S1 & From EOS, Eq.\ S2 \\
\hline
THz pump  & 6.5 MV/cm & (6.2$\pm$1.2) MV/cm \\
\hline
\end{tabular}
\caption{Comparison between the THz electric field strengths calculated from Eq.\ S1 (by measuring THz energy, temporal profile and spot size) and from Eq.\ S2 (by measuring the electro-optic modulation). The uncertainty has been estimated based on the uncertainty on transmittance of the attenuators.}
\label{tableS1}
\end{table}

Measurements at different temperatures (Fig.\ 4 of the main manuscript) were performed by using a liquid nitrogen (LN) optical cryostat in the range 77 K - 300 K and by using a sample heater in the range 325 K - 500 K. The optical cryostat was equipped with two TPX windows to transmit both the THz pump and the NIR probe (the sample was in vacuum inside the cryostat). In order to keep the optical transport unchanged for the measurements with the sample heater, a TPX window was put in front of the sample. The TPX window transmits $\sim 90\%$ of the THz electric field.

\section*{Supplementary Note 1. THz spectroscopy of V$_2$O$_3$ thin film}

We measured the temperature evolution of the THz optical conductivity across the metal to insulator transition (MIT) by means of Fourier transform infrared spectroscopy in the THz range~\cite{ref4}. Transmittance of V$_2$O$_3$ thin film on Al$_2$O$_3$ substrate used in the THz pump experiment was measured in the range 77 K - 325 K with a FTIR interferometer (Vertex 80V) coupled to a continuous flow optical cryostat (Janis ST-100-FTIR) (Fig.\ S2 {\bf a}). We observe a decrease of the transmitted THz intensity of a factor $\sim 10$ from 77 K (insulating phase) to 200 K (metallic phase) due to the increase in optical conductivity across the MIT ($T_{MIT}\sim150$ K).

Optical conductivity in the THz range has been determined from the THz intensity transmitted through the film on substrate normalized to that of the bare substrate according to the Tinkham formula ~\cite{ref4,ref5}. The real part of the THz optical conductivity $\sigma_1$ is shown in Fig.\ S2{\bf b}. $\sigma_1$ vs frequency is constant in the THz range, while the imaginary part $\sigma_2$ is negligible as reported in Ref.s ~\cite{ref4, ref6}. The real part $\sigma_1$ at 4 THz as a function of temperature is shown in the inset of Fig.\ S2{\bf b}.
\begin{figure}[h!]
\centering
\includegraphics[scale=0.60]{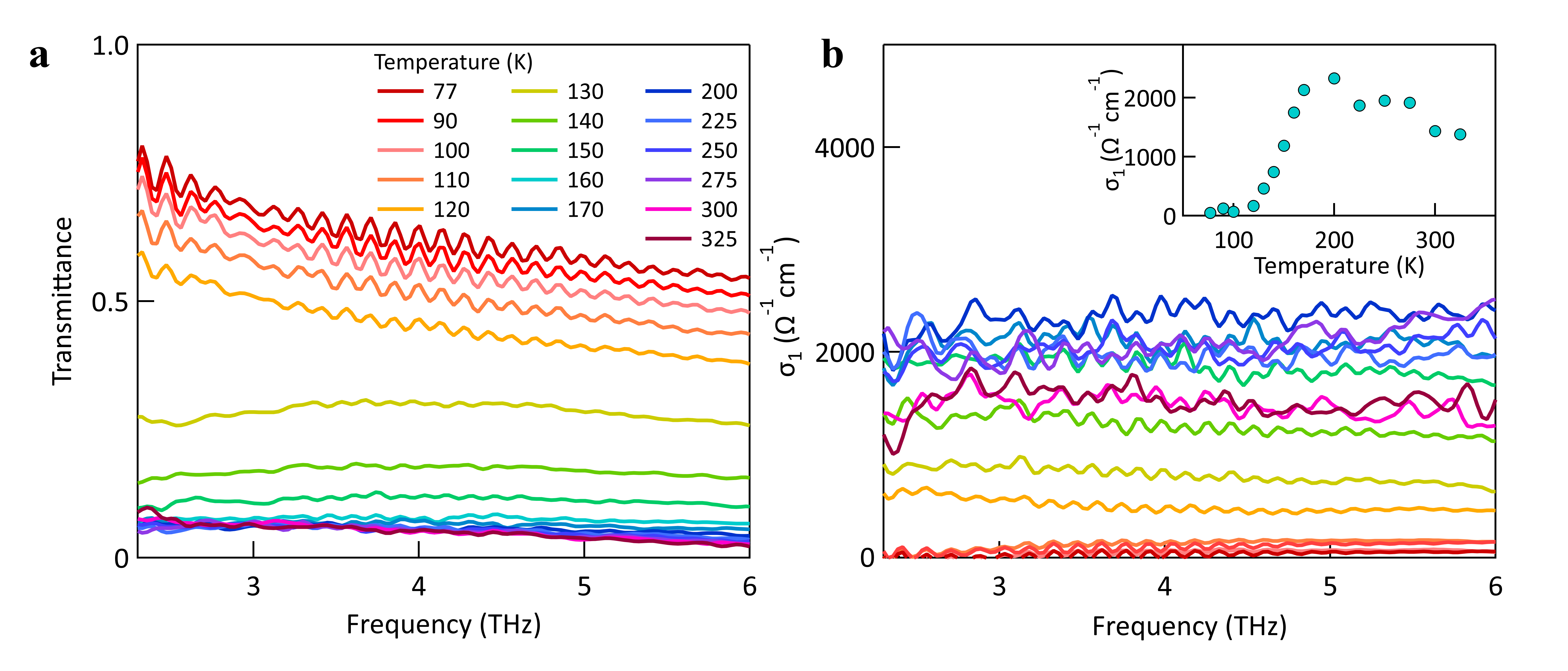}
\caption{\textbf{Optical response of V$_2$O$_3$ thin film in the THz range}. {\bf a}) Temperature evolution of transmittance of V$_2$O$_3$ thin film on Al$_2$O$_3$ substrate in the THz range 77 K - 325 K. {\bf b}) Temperature evolution of $\sigma_1$ in the THz range across the MIT. Inset: $\sigma_1$ at 4 THz as a function of temperature.}
\end{figure}

\section*{Supplementary Note 2. Contribution from the Al$_2$O$_3$ substrate in the observed THz pump-optical probe dynamics}

To further demonstrate that the coherent oscillations at twice the frequency of the THz pump are uniquely related to the response of the V$_2$O$_3$ thin film, we compared the THz-pump-induced optical transmission modulation of V$_2$O$_3$ on Al$_2$O$_3$ substrate to the one of the bare Al$_2$O$_3$ substrate in the same experimental condition (Fig.\ S3). As observed, the Al$_2$O$_3$ shows no transmission modulation upon THz irradiation. In the inset for Fig.\ S3, we report the temporal derivative of the THz induced dynamics for both V$_2$O$_3$ on Al$_2$O$_3$ and bare Al$_2$O$_3$. The dynamics of Al$_2$O$_3$ shows neither oscillations nor THz-induced transmission modulation.
\begin{figure}[h!]
\centering
\includegraphics[scale=0.65]{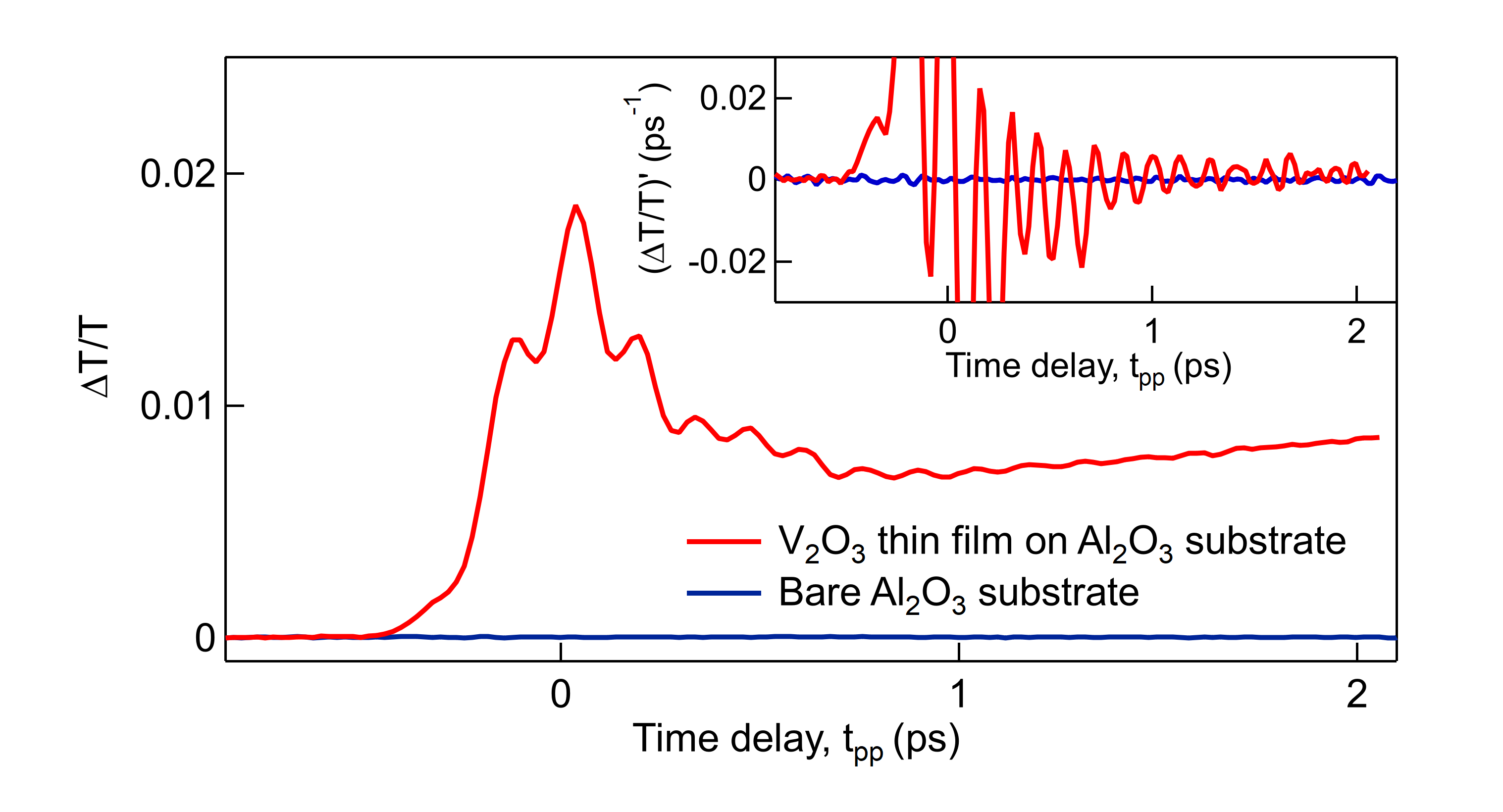}
\caption{\textbf{Contribution from the Al$_2$O$_3$ substrate}. THz-induced optical transmission modulation of V$_2$O$_3$ film on R-plane Al$_2$O$_3$ (red curve) and of bare R-plane Al$_2$O$_3$ substrate (blue curve). Inset: time derivative of the dynamics.}
\end{figure}

\section*{Supplementary Note 3. Phononic contributions to Raman spectrum of V$_2$O$_3$}
Fig.\ S4{\bf a} shows the Raman spectrum of V$_2$O$_3$ at room temperature (metallic phase). The Raman spectrum consists of three peaks that can be assigned to vibrational modes with A$_{1g}$ and E$_g$ symmetries~\cite{ref7,ref8}. The peaks from the A$_{1g}$ and lowest frequency E$_g$ modes are partially overlapped~\cite{ref7,ref8}. In order to evaluate the contribution of the E$_g$ phonon modes, we have fitted the Raman intensity, see Fig.\ S4{\bf a} through a multi-component model for the A$_{1g}$ peak and the two E$_g$ peaks, as described by A.\ Okamoto et al.\ (Ref.\ ~\cite{ref8}). The A$_{1g}$ phonon has an asymmetric line shape characterized by the Fano resonance effect due to the interaction of the phonon with the continuum of electronic states~\cite{ref8}. The E$_g$ modes have Lorentzian (symmetric) line shapes.
\begin{figure}[h!]
\centering
\includegraphics[scale=0.6]{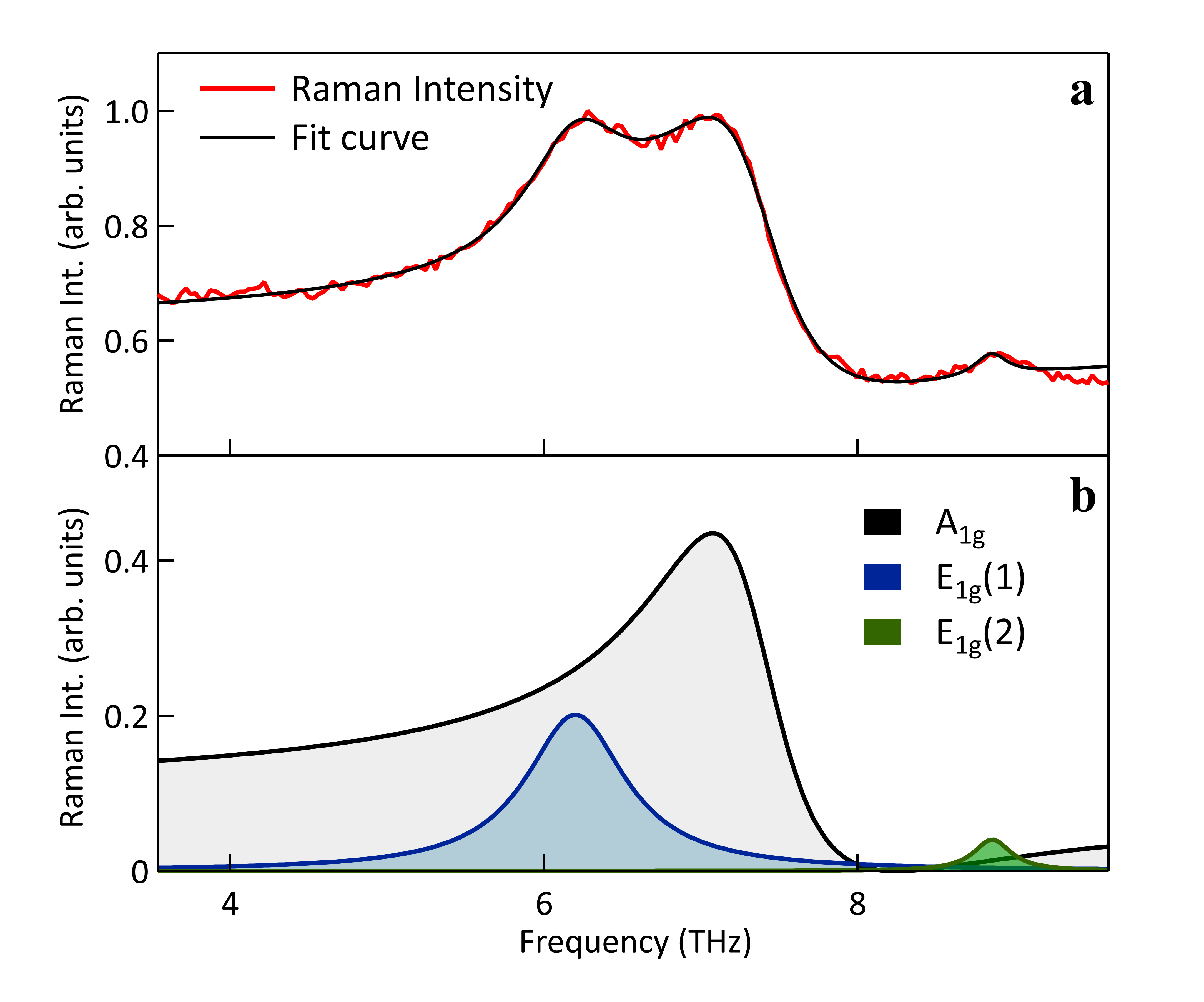}
\caption{\textbf{Raman spectroscopy of V$_2$O$_3$ thin film}. {\bf a}) Raman spectrum of V$_2$O$_3$ thin film at 300 K (metallic phase) and fit using Eq.\ S3. {\bf b}) A$_{1g}$, E$_g$(1) and E$_g$(1) phonon line shapes from the single components of the fitting curve.}
\end{figure}

According to A.\ Okamoto et al.\ (Ref.\ ~\cite{ref8}), the Raman spectrum can be modeled as:
\begin{equation}
I(\omega)= I_0 \frac{\left( q+ \frac{\omega-\omega_0}{\Gamma_0} \right)^2}{1+\left( \frac{\omega-\omega_0}{\Gamma_0} \right)^2}+\frac{I_1}{(\omega-\omega_1)^2+\left(\frac{\Gamma_1}{2} \right)^2}+\frac{I_2}{(\omega-\omega_2)^2+\left(\frac{\Gamma_2}{2} \right)^2}+ c. 
\label{eqS3}
\end{equation}
The first term represents the Fano line shape of the A$_{1g}$ phonon mode~\cite{ref8}, where $I_0$ is the Raman intensity, $\Gamma_0$ is the measured line width and q$^{-1}$ is the Fano coupling coefficient. $\omega_0$ is the resonance frequency. The second and the third terms in Eq.\ S3 are the Lorentzian functions describing the Raman intensity of E$_g$ phonons, which are peaked at frequencies $\omega_1$ and $\omega_2$ with strengths $I_1$ and $I_2$ and widths $\Gamma_1$ and $\Gamma_2$, respectively. The offset variable $c$ accounts for the Raman intensity background. We used Eq.\ S3 to fit the experimental data and to extract the phonon parameters. The fit model, depicted in Fig.\ S5{\bf a} (black curve) captures the experimental data well. The individual phonon components are shown in Fig.\ S4{\bf b}. Fit parameters are reported in Table S2.\\

Due to the asymmetry of the Fano line shape, the A$_{1g}$ phonon mode is peaked at 236 cm$^{-1}$ (7.1 THz), which matches the frequency of the oscillations observed in the THz pump-probe dynamics (see main manuscript).\\
\begin{table}[h]
\centering
\renewcommand{\arraystretch}{1.5}
\begin{tabular}{| c | c | c | c | c | c | c | c | c | c | c | }
\hline
\multicolumn{4}{|c|}{A$_{1g}$} & \multicolumn{3}{c|}{E$_{g}$(1)} & \multicolumn{3}{c|}{E$_{1g}$(2)} & \\
\hline
$I_0$ & $\omega_0$ (cm$^{-1}$) & $\Gamma_0$ (cm$^{-1}$) & q & $I_1$(cm$^{-2}$) & $\omega_1$ (cm$^{-1}$) & $\Gamma_1$ (cm$^{-1}$) & $I_2$(cm$^{-2}$) & $\omega_2$ (cm$^{-1}$) & $\Gamma_2$ (cm$^{-1}$) & c  \\
\hline
0.0934 & 244 & 16 & -1.91 & 33.5 & 206 & 166 & 0.93 & 295 & 22.9 & 0.519 \\
\hline
\end{tabular}
\caption{Phonon parameters by fit procedure using Eq.\ S3.}
\label{tableS2}
\end{table}

To show that the observed oscillations are related to the A$_{1g}$ phonon mode we analyze the symmetry of the phonon modes. A$_{1g}$ and E$_g$ phonon modes are, respectively, fully symmetric and non-fully symmetric modes. Their Raman tensors in the $x$, $y$, and $z$ coordinates (where $z$ corresponds to the $c$-axis) can be expressed as:

$$ A_{1g}^{xyz}= \begin{bmatrix}
    a	& 0 & 0 \\
    0	& a & 0 \\
    0	& 0 & b \\
\end{bmatrix}; \hspace{1cm}
E_g^{xyz}= \begin{bmatrix}
    c	& 0 & 0 \\
    0	& -c & d \\
    0	& d & 0 \\
\end{bmatrix}. $$
Due to the symmetry of the Raman tensors, the transient changes of transmittance of the sample are only related to the fully-symmetric A$_{1g}$ mode while changes in optical polarization arises from the off-diagonal elements of the E$_g$ mode~\cite{ref9}. In principle, the observed oscillations in the dynamics are induced by the A$_{1g}$ mode. However, a potential contribution of the E$_g$ mode may be observed because the V$_2$O$_3$ film is $c$-axis oriented at $\theta=57.62^{\circ}$ with respect to its surface~\cite{ref10}. The contribution of E$_g$ to the measured transmission modulation can be quantitatively evaluated by calculating the Raman tensors in new coordinate systems $x'$, $y'$, and $z'$, in which the $z$ axis is rotated by $\theta$ around $x$ and $\phi$ rotation describes the azimuthal orientation of the crystal on the substrate:
$$ A_{1g}^{x'y'z'}=  R_x^T R_z^T A_{1g}^{xyz} R_z R_x= \begin{bmatrix}
    a	& 0 & 0 \\
    0	& a\cos^2 \theta + b \sin^2 \theta & (-a+b)\cos\theta\sin\theta \\
    0	& (-a+b)\cos\theta\sin\theta & b\cos^2 \theta + a \sin^2 \theta \\
\end{bmatrix},$$
and
\begin{gather*}
E_{g}^{x'y'z'}=  R_x^T R_z^T E_{g}^{xyz} R_z R_x= \\
\begin{bmatrix}
c \cos 2\phi		& (-2c \cos\theta \cos\phi+ d \sin\theta)\sin\phi  & (d \cos\theta+2c \cos\phi \sin\theta)\sin\phi \\
(-2c \cos\theta \cos\phi+ d \sin\theta)\sin\phi	& \cos\theta(-c\cos\theta \cos 2\phi+2 d \cos\phi \sin\theta) & d \cos2\theta \cos\phi+\frac{1}{2} c \cos2\phi \sin2\theta \\
 (d \cos\theta+2c \cos\phi \sin\theta)\sin\phi	& d \cos2\theta \cos\phi+\frac{1}{2} c \cos2\phi \sin2\theta & -\sin \theta (2d\cos\theta\cos\phi + c\cos2\phi\sin\theta) \\
\end{bmatrix}.
\end{gather*}
$R_x (\theta)$ and $R_z (\phi)$ are rotation matrices and $R_x^T (\theta)$ and $R_z^T (\phi)$ their transposes. Since the V$_2$O$_3$ thin film is epitaxial, crystal domains are oriented on the substrate with three different $\phi$ at 120$^{\circ}$ rotation. $A_{1g}^{x'y'z'}$ is $phi$ independent while averaging $E_g^{x'y'z'}$ over $\phi=\{0,2/3\pi,4/3\pi \}$ each term goes to zero. Therefore we can conclude that the main contributors to the observed oscillations induced by THz displacive excitation are related to the A$_{1g}$ mode.

As a further cross-check, we have performed the experiment on a V$_2$O$_3$ film deposited on the c-cut (0001) Al$_2$O$_3$ substrate. For this sample, the $c$-axis of V$_2$O$_3$ crystal is perpendicular to the surface as well as along the propagation direction of the THz pump and NIR probe pulses. This crystal orientation results in a fully-symmetric A$_{1g}$ Raman tensor with only diagonal terms, which induce isotropic variations of the transmission, while the E$_g$ modes contribution cancel out by symmetry~\cite{ref9}. Thus, we expect only the A$_{1g}$ phonon to be involved in the THz displacive excitation.

For the experiment, we deposited a $\sim$80 nm V$_2$O$_3$ film on c-cut Al$_2$O$_3$ by pulsed laser deposition (PLD). The epitaxial growth conditions are achieved with an oxygen partial pressure of $10^{-6}$ mbar while the temperature of the substrate is kept at 720 $^{\circ}$C. The V$_2$O$_3$ polycrystalline target was ablated using a solid-state laser ($\lambda$ = 266 nm). In Fig.\ S5{\bf a}, we show the THz pump-NIR probe dynamics of the V$_2$O$_3$ film on c-cut Al$_2$O$_3$ measured in the same experimental conditions of the one on R-plane Al$_2$O$_3$ (Fig.\ 1{\bf e} of the main manuscript). The THz pump-induced transmission modulation shows coherent oscillations at frequencies higher than the spectral contents of THz pump, see inset of Fig.\ S5{\bf a}, as observed for the V$_2$O$_3$ film on R-plane Al$_2$O$_3$. The observed oscillations peak at the frequency of the A$_{1g}$ phonon mode measured by Raman spectroscopy as shown in Fig.\ S5{\bf b}-{\bf c}.
\begin{figure}[h!]
\centering
\includegraphics[scale=0.60]{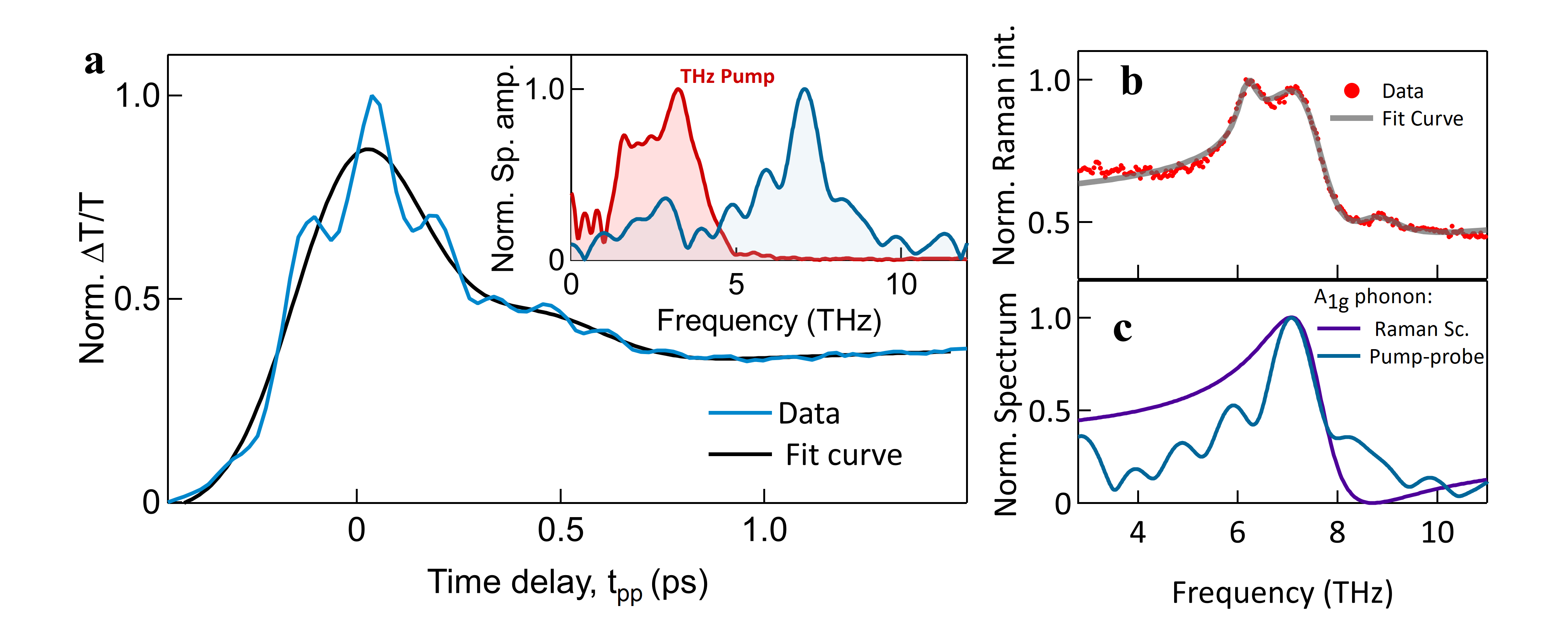}
\caption{\textbf{THz pump-induced transmission modulation dynamics in a V$_2$O$_3$ film on deposited on the c-cut Al$_2$O$_3$ substrate}. {\bf a}) Experimental data (blue solid curve) and fit (black solid curve) of the slowly-varying background using Eq.\ S4. Inset: Fourier transform of the oscillatory component for $t_{pp}> 4.6$ ps. {\bf b}) Raman spectrum and fitting curve using Eq.\ S3. {\bf c}) Raman spectrum of the A$_{1g}$ phonon mode from fitting procedure (Eq.\ S3) and spectrum of the pump-probe dynamics.}
\end{figure}

\section*{Supplementary Note 4. Fit procedure to extract the oscillatory component}

As discussed in the manuscript (see Fig.\ 1 {\bf e}-{\bf f}), coherent oscillations have been extracted by subtraction of a smooth background determined by the following function:
\begin{equation}
f(x) = A \mathcal{G}(x,t_0,w_1)+B \left[ erf\left( -\frac{(x-t_0)}{w_2} \right) +1 \right] e^{-\frac{(x-t_0)}{w_3}} + C \left[ erf\left( -\frac{(x-t_0)}{w_4} \right) +1 \right] e^{-\frac{(x-t_0)}{w_5}}+ D \mathcal{G}(x,t_1,w_6),
\end{equation}
where $\mathcal{G}(x,t,w)=\frac{1}{w\sqrt{2\pi}}e^{-\frac{1}{2}\left( \frac{x-t}{w} \right)^2}$ is the normalized Gaussian function and $erf(x)$ is the error function. The Eq.\ S4 consists of the sum of four terms. The first term is a Gaussian function that phenomenologically captures the THz excitation. The second term is a step-like function to reproduce the slowly-varying background of the thermalization dynamics after the THz excitation. The third and fourth terms are small corrections. Specifically the fourth term represents a Gaussian line shape in the dynamics at 0.46 ps that may originate from a second reflection THz pulse as observed in the EOS time-trace, see Fig.\ S1{\bf b}.

The curve fitting results by using Eq.\ S4 is shown in Fig.\ S6{\bf a}. The fit parameters are reported in Table S3. The four terms of Eq.\ S4 are plotted separately in Fig.\ S6{\bf b}.
\begin{figure}[h!]
\centering
\includegraphics[scale=0.8]{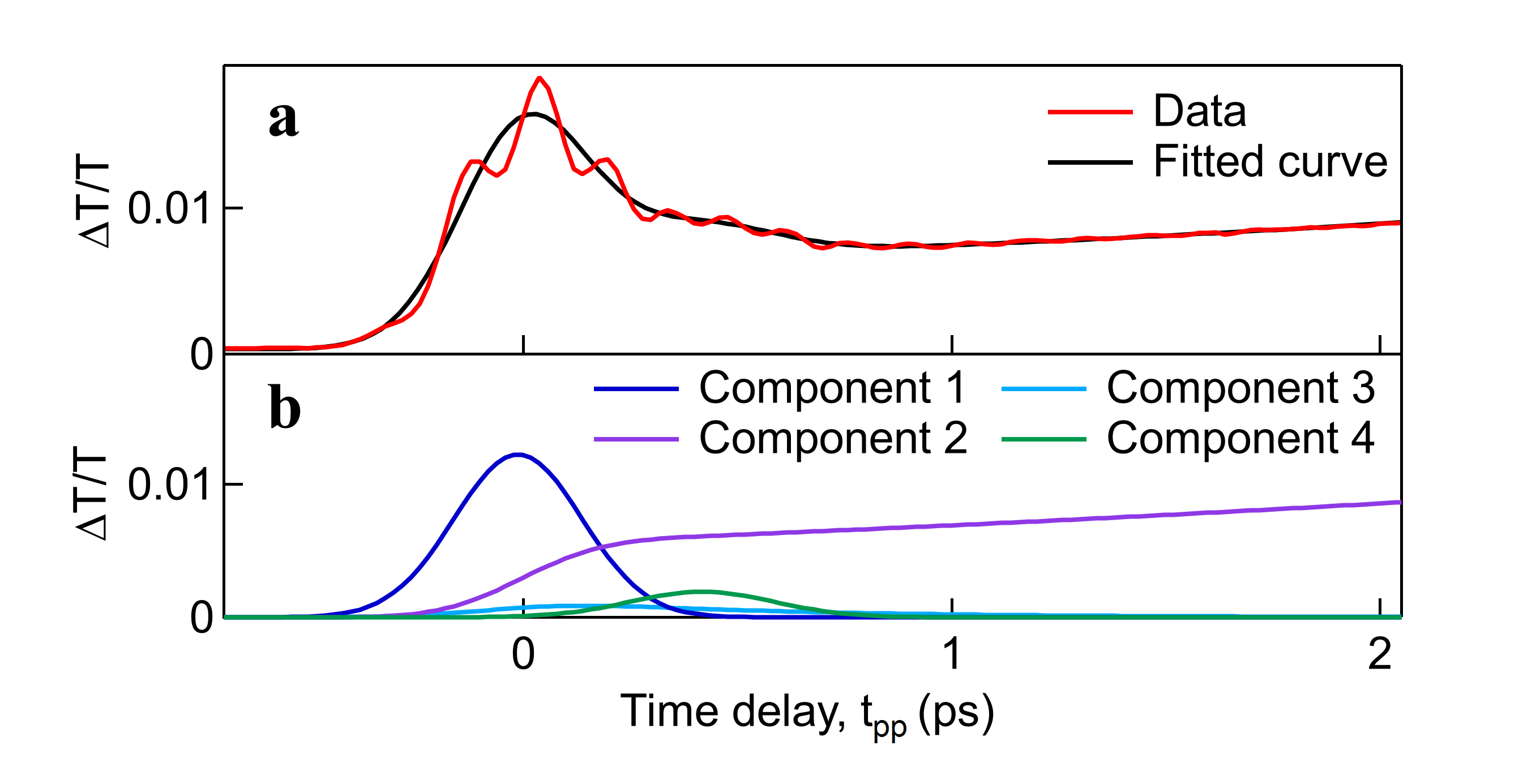}
\caption{\textbf{Background fitting}. {\bf a}) Experimental data and curve fitting by using Eq.\ S4. {\bf b}) The four terms of Eq.\ S4 plotted separately.}
\end{figure}
\begin{table}[h]
\centering
\renewcommand{\arraystretch}{1.7}
\begin{tabular}{| c | c | c | c | c | c | c | c | c | c | c | c |}
\hline
A & $t_0$ (ps) & $w_1$ (ps) & B & $w_2$ (ps) & $w_3$ (ps) & C & $w_4$ & $w_5$ (ps) & D & $t_1$ (ps) & $w_6$ (ps)\\
\hline
0.0046 & -0.013 & 0.15 & 0.028 & -0.21 & -4.7 & 0.00070 & -0.23 & 0.54 & 0.00080 & 0.46 & 0.17 \\
\hline
\end{tabular}
\caption{Fit parameters obtained using Eq.\ S4}
\label{tableS1}
\end{table}

Fig.\ S7 shows the oscillatory components of the THz-induced transmission modulation at 77 K (insulating phase) and at 240 K (metallic phase) of the experimental curves in Fig.\ 4c of the manuscript, after subtracting the slowly-varying background by using Eq.\ S4. As discussed in the main manuscript, the A$_{1g}$ phonon mode generated at twice the frequency of the THz pump pulse is coherently excited by a displacive force arising from the THz excitation of intraband electronic transitions that are strongly coupled to the lattice mode. This mechanism can be seen as a low-frequency counterpart of the DECP for opaque materials described by Zeiger \textit{et al.}~\cite{ref11} and generalized by Merlin\textit{et al.}~\cite{ref12}, who discussed the real and imaginary parts of the Raman tensor involved in the coherent phonon generation. The real part, dominant in transparent solids, leads to an impulsive force $F(t)=|E(t)|^2$ on the crystal lattice, where $E(t)$ is the electric field. The imaginary part, dominant in opaque materials, induces a displacive force $F(t)=\int_{-\infty}^t |E(t')|^2 dt'$. For the metallic phase of V$_2$O$_3$, the THz generation mechanism is governed by the imaginary part of the Raman tensor, whose intensity - peaked at the phonon frequency - is enhanced by the coupling with the electronic system, as discussed in Ref.\ ~\cite{ref8}. During the metal-insulator transition, the intensity of the Raman tensor decreases by more than a factor 5 (see Ref.\ ~\cite{ref13}) explaining the reduction in amplitude of the phononic oscillations induced by THz excitation (see Fig.\ S7). The decrease in the intensity of the Raman tensor across the metal-to-insulator transition is primarily described by the absence of the intraband electronic transitions~\cite{ref8}. However, an accurate comparison of the oscillation amplitude of the two phases is difficult because the metal-to-insulator transition is also accompanied by a structural transition leading to a change in the phonon spectrum in which the A$_{1g}$ mode transforms into a frequency-upshifted A${_g}$ mode~\cite{ref7}.

From theory, the two driving forces are phase-shifted by $\pi/2$ with respect to each other~\cite{ref12}, which is reflected in the data in Fig.\ S7 (dashed lines), suggesting that for the insulating phase the regime is impulsive and for the metallic phase it is displacive. However, the intense THz field excitation of the insulating phase triggers an ultrafast metallization~\cite{ref14}, for which the discrimination between the impulsive and displacive regime is elusive and goes beyond the scope of this article. Nonetheless, we can still conclude that the regime in the metallic phase is purely dispersive as discussed in the main text.
\begin{figure}[h!]
\centering
\includegraphics[scale=0.65]{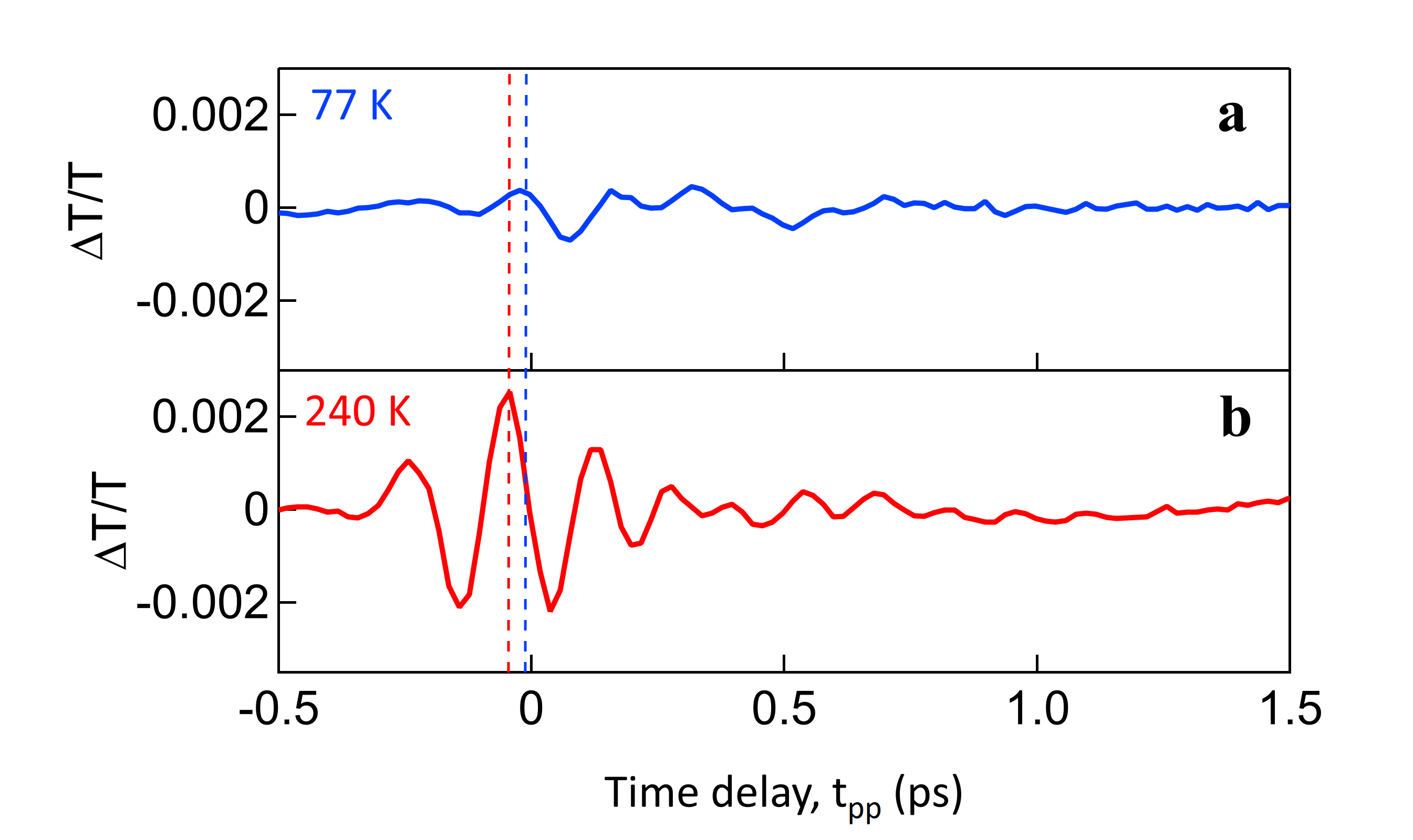}
\caption{\textbf{Coherent oscillations after subtracting the background} at ({\bf a}) 77 K and at ({\bf b}) 240 K.}
\end{figure}

\end{document}